\documentclass[journal]{IEEEtran}

\usepackage{graphicx}
\usepackage{hyperref}
\graphicspath{{./figures/}}
\usepackage{epstopdf}
\DeclareGraphicsExtensions{.eps}
\usepackage{amsmath}
\usepackage{amssymb}
\usepackage{tabularx}
\usepackage{tabu}
\usepackage{multirow}
\usepackage{hhline}% http://ctan.org/pkg/hhline
\usepackage{pbox}
\usepackage{url}
\usepackage{algorithm}
\usepackage{algpseudocode}
\usepackage{xcolor}
\usepackage{colortbl}
\begin{document}

\title{Can the Multi-Incoming Smart Meter Compressed Streams be Re-Compressed? }

\author{Sharif Abuadbba, Ayman Ibaida, Ibrahim Khalil, Naveen Chilamkurti, Surya Nepal and Xinghuo Yu
       
\IEEEcompsocitemizethanks{\IEEEcompsocthanksitem Sharif Abuadbba and Surya Nepal are with Data61, CSIRO  and Cyber Security CRC, Australia. e-mail: \{sharif.abuadbba, surya.nepal\}@data61.csiro.au\protect\\
Assoc/Prof. Ibrahim Khalil and Prof. Xinghuo Yu are with RMIT University, Australia. e-mail: \{ibrahim.khalil,x.yu\}@rmit.edu.au\protect\\ Dr. Ayman Ibaida is with Victoria University, Australia. e-mail:  ayman.ibaida@vu.edu.au\protect\\ Assoc/Prof. Naveen Chilamkurti is with Latrobe University, Australia. email: naveen.chilamkurti@latrobe.edu.au.
}% <-this % stops a space
\thanks{}
}
%\author{Sharif~Abuadbba,~\IEEEmembership{Student Member,~IEEE}
    %    and~I~Khalil,~\IEEEmembership{Member,~IEEE}
       % <-this % stops a space
%\IEEEcompsocitemizethanks{\IEEEcompsocthanksitem Alsharif Abuadbba is with the School of Computer Science and Information Technology, RMIT University, Melbourne, VIC, Australia, 3001.\protect\\
%%%\IEEEcompsocitemizethanks{\IEEEcompsocthanksitem Alsharif Abuadbba and Ibrahim Khalil are with the School of Computer Science and Information Technology, RMIT University, Melbourne, VIC, Australia, 3001.\protect\\
%%%E-mail: alsharif.abuadbba@rmit.edu.au and ibrahim.khalil@rmit.edu.au}% <-this % stops a space
%\thanks{}
%}

%IEEE TRANSACTIONS ON SMART GRID 2016
\markboth{}%
{Shell \MakeLowercase{\textit{et al.}}: Bare Demo of IEEEtran.cls for Computer Society Journals}

\newcommand{\sharif}[1]{\textcolor{blue}{Sharif: #1}}

\IEEEtitleabstractindextext{%
\begin{abstract}
Smart meters have currently attracted attention because of their high efficiency and throughput performance. They transmit a massive volume of continuously collected waveform readings (e.g. monitoring). Although many compression models are proposed, the unexpected size of these compressed streams required endless storage and management space which poses a unique challenge. Therefore, this paper explores the question of  can the compressed smart meter readings  be  re-compressed? We first investigate the applicability of re-applying general   compression   algorithms   directly   on   compressed streams. The results were poor due to the lack of redundancy.  We  further propose  a  novel  technique  to enhance  the  theoretical  entropy  and  exploit  that  to  re-compress.  This is successfully achieved by using unsupervised  learning as a similarity measurement to cluster the compressed streams into subgroups. The streams in every subgroup have been interleaved,   followed by the first derivative to minimize the values and increase the redundancy. After that, two rotation steps have been applied to rearrange the readings in a more consecutive format before applying a developed dynamic run length. Finally, entropy coding is performed. Both mathematical and empirical experiments proved the significant improvement of the compressed streams entropy (i.e. almost reduced by half) and the resultant compression ratio (i.e. up to 50\%). 	
\end{abstract}

}

\maketitle

\IEEEdisplaynontitleabstractindextext

\IEEEpeerreviewmaketitle

\IEEEraisesectionheading{}

\section{Introduction}\label{sec:introduction}

Due to the lack of outage management, automation, poor real-time analysis and deficiency of the classic power grid of the past century, smart meters are currently being investigated around the world. They automatically collect periodic waveform readings every second (e.g. power consumption of the premise) and transmit them to operational centers (e.g. cloud servers) using various techniques \cite{smartgrid:gungor2010opportunities}. The International Council of Large Electric Systems (CIGRE) recent survey \cite{CIGREwebsite} highlights that there are more than twelve (12) key applications (i.e., use cases) that  might be accomplished from the distributed smart meters. On top of the list are load prediction, automatic metering services, and energy feedback. 
% Therefore, it is expected to see an exponential increase in the volume of collected reading from
% small meter devices to operational centres. 
Only one study over US Western states produced  100 Terabyte of smart meters data over 3 months with 220+ Gigabyte per day \cite{smartgrid:gungor2011smart}.

\textbf{Existing Landscape}: The proposed compression methods for smart meter waveforms readings can be categorized into two groups - lossy and lossless \cite{cp:stateofart:tcheou2014}. Lossy compression depends on  losing some information while preserving the main features of the waveforms signal. Consequently, the decompressed signal is somewhat dissimilar to the original. This type of compression was acceptable in the classical grid model, and so lots of work has been done in this direction which can be grouped into transformation techniques~\cite{cp:dwtsg:ning2011wavelet,cp:liftingwp:tse2012real},  parametric coding \cite{cp:damsin:tcheou2007optimum} and mixed \cite{cp:harmonic:ribeiro2007novel}. This is due to its ability to achieve  a higher compression ratio while losing some data. However, lossy compression is recently not recommended for the two following reasons. (1)  The smart meters collected readings  are potentially used in billings and other purposes. (2) To preserve the privacy and authenticity of the transmitted readings, recent models are utilizing watermarking to conceal the private information randomly inside these readings \cite{stego:abuadbba2014wavelet}. 

%Therefore, losing any bit of these readings is unacceptable.

In contrast, lossless compression is obligated to recover the same waveform signal as the original with zero loss. Due to these restrictions, a few works have been proposed under this category such as in \cite{cp:less:singularval:2015desouza,cp:less:transgolomb:2016tale, tripathi2018efficient}. This includes our recent lossless compression work \cite{cp:abuadbba2017gaussian} where the compression ratio has been doubled from existing techniques. However, according to \cite{cp:stateofart:tcheou2014}, this path is far from being as mature as image, voice and video lossless compression.  

\textbf{Limitations:} Most of the existing research has been done to compress the transmitted streams from the premises to intermediate gateways or  operation centers (i.e. public or private cloud servers). However, little attention has been paid to how to handle the multi-incoming compressed streams after arriving at intermediate hubs or  cloud level. Especially, due to some regulations these streams should be stored for a certain number of years. This means an exponential increase in storage space cost and hardening of data management which we are compelled to addressed.
% \begin{figure*}[!t] %proposedarchiticture
% 	\centerline
% 	{\includegraphics[scale=0.35]{proposedarchitecture.eps}}
% 	\caption{The main scenario of our proposed technique where the multi-incoming compressed streams are categorized by their  similarity features  followed by interleaving and lossless size reduction.} 
% 	\label{fig:proposedarceticture}
% 	\hfil
% \end{figure*} 

%One would suggests recompressing these streams. However, this is not a straightforward task. The reason beyond that is the difficulty of re-compressing the compressed streams especially after lossless compression. This is due to  that  the mathematical entropy (i.e. the minimum number of bits required to represent a value after compression) has been already exploited during the initial compression  by eliminating existing redundancy. 

Therefore, this work is dedicated to address this challenge by investigating the answers to the following  {\bf research question (RQ):} \vspace{0.1cm}
\noindent\textit{\textbf{ Can the multi-incoming smart meter compressed streams be re-compressed?}}  To answer RQ, we have made the following contributions: 
 \begin{itemize}
     \item We first investigate the applicability of re-applying general compression algorithms directly on compressed streams. The results were poor due to the lack of  redundancy which is exploited in the first compression stage.
     \item To address the shortcoming of direct application of compression algorithms, we  propose a novel technique to enhance the theoretical entropy (i.e. the minimum number of bits required to represent a value after compression) and exploit that to re-compress. To the best of our knowledge, this is the first elaborated work on compressing the already compressed streams.
 \end{itemize}
 
% Fig. \ref{fig:modelflow} demonstrates an overview of our proposed model where  unsupervised learning technique (i.e., K-Means clustering) is used  as a similarity measurement to classify the compressed streams into subgroups at intermediate hubs or cloud level. The streams in every subgroup have been interleaved  followed by the first derivative to reduce the values and increase the redundancy. After that, two rotation steps,  including Burrow Wheeler Transform (BWT) and Move-To-Front (MTF) have been applied to rearrange the readings in a more consecutive format. Finally, Dynamic RLE and entropy coding are performed. We proved that it is possible to re-compress the already compressed streams up to $50\%$ of their size. %To the best of our knowledge, there is no other work that tackles this issue in the field of smart meter readings.

\section{Related Work}\label{sec:relatedWork}

Most of the research has been conducted on waveforms gathered readings targeted at lossy compression. This is due to that (i) the samples were not directly transmitted and utilized for crucial purposes such as real-time diagnoses and billings in the classical grid system, and (2) the transformation techniques such as wavelet transform are more efficient in representing waveform signals in few values (i.e. with losing some bits from every sample). The lossy compression studies can be classified based on the used techniques into transformation, parametric coding and mixed.

Transformation models include the work of  Santoso et al. \cite{cp:dwt:santoso1997power}, where discrete wavelet decomposition has been employed to identify most of the signal energy in low-frequency coefficients (i.e. using dbX) while neglecting other coefficients during compression. Further work has been conducted using various wavelet families  such as  B-Spline \cite{cp:bspline:meher2004integrated} and Sluntlet \cite{cp:swt:panda2002data}. Secondly, Parametric coding models such as the work of Michel et al. \cite{cp:damsin:tcheou2007optimum} where they utilized damped sinusoids models to elicit the main features of the signal before compression. Finally, mixed transformation and parametric models include the work proposed by Moises et al. \cite{cp:harmonic:ribeiro2007novel} where they employed fundamental harmonic and transient coding together. 

In contrary, less work has been pursuing the lossless compression due to the imposed constraints and the nature of waveforms readings. The lossless compression can be categorized based on the technique utilized into the dictionary, entropy and mixed based models.

Dictionary-based techniques rely mainly  on general compressors (e.g. GZIP, ZIP and LZO) where a dictionary is constructed, and more frequent tokens will be represented in fewer bits. In contrast,  more bits  are assigned to less frequent samples. For instance, Omer and Dogan \cite{cp:less:lempelziv:gerek2008} employed Lempel-Ziv to compress a stream of waveforms readings. The accomplished compression ratio was 2.5:1 bin-to-bin. However, dictionary algorithms are primarily designed for letters (e.g. English characters) where the number of choices is limited. This is unsuitable for  waveforms signals due to their floating-point nature. This means every real number has thousands of forms because of its floating values.

Entropy-based techniques are essentially statistical models designed based on screening the probability of every token within a stream and assigning less number of bits for higher probability and vice-versa. For instance, K. Jan et al. \cite{cp:less:ac:bzip2:kraus2009loooseless}  employed arithmetic coding to replace the input tokens with a single floating-point value. The accomplished compression ratio was 2.6:1. Z. Dahai et al.  \cite{cp:less:Huff:zhang2009new} also introduced a model that enhances Huffman coding by preprocessing the data utilizing higher order delta modulation. The enhancement was from 1.7 to 2.3:1. Moreover, J. Tate \cite{cp:less:transgolomb:2016tale}  proposed a model that utilizes Golomb-Rice coding after preprocessing the samples with several methods such as frequency compensated difference. The accomplished compression ratio was 2.8:1. Finally, in our recent model \cite{cp:abuadbba2017gaussian}, we used Gaussian approximation followed by entropy coding techniques and the achieved compression ratio was 3.8:1. 

Mixed techniques are more sophisticated algorithms using both dictionary and statistical mechanisms. This permits exploiting both the frequency of repetition and its probability within a stream of samples. For instance, K. Jan et al. \cite{cp:less:predictionlzma:kraus2012} proposed a model that enhances LZMA algorithm to minimize the redundancy in waveforms readings. This is achieved by utilizing  prediction techniques based on differential encoding after optimizing the interval selection. The accomplished compression was 2.6:1. K. Jan and T. Tomas \cite{cp:less:ac:bzip2:kraus2009loooseless} also proposed a model that enhances BZIP2 by employing an efficient block sorting Burrows-Wheeler algorithm and delta modulation. The achieved compression ratio was 2.9:1. 

All the above models have been conducted on original stream readings and so they exploited the existing high redundancy probabilities among them. We are not aware of current  lossless compression work that targeted  compressed streams. 

\section{Applying General Compressors Directly}
In this section, we investigate the applicability of applying general compressors to re-compress already compressed streams. Specifically, we explore the following ideas:
%\begin{enumerate}

    (A) Can we re-compress a single compressed stream?
    
    (B) Can we re-compress multiple compressed streams together?
%\end{enumerate}

\begin{figure}[!h] %proposedarchiticture
	\centerline
	{\includegraphics[scale=0.43]{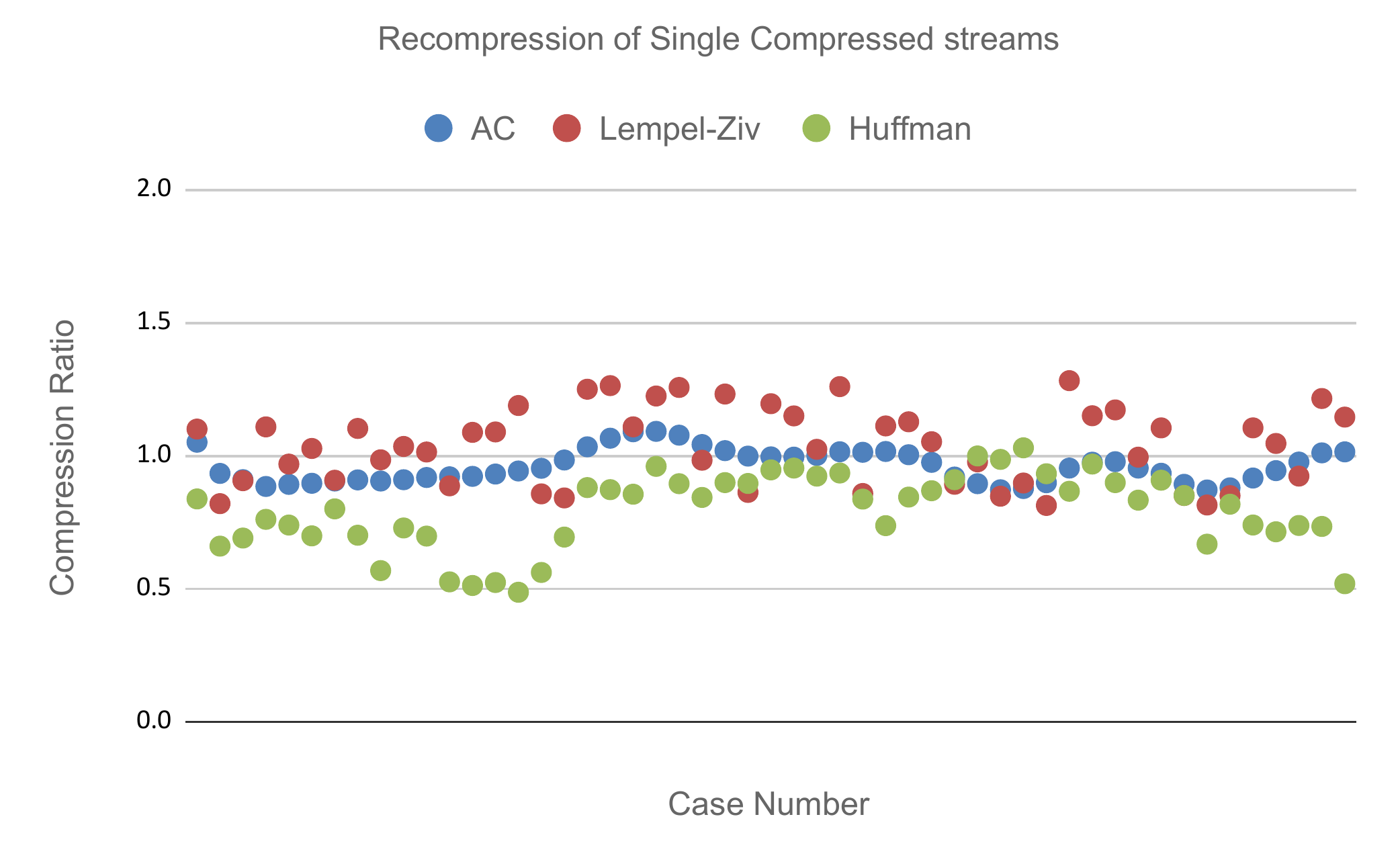}}
	\caption{Compression ratio after re-compressing single incoming streams directly using various well-known lossless compressors.} 
	\label{fig:singlestreamCR}
	\hfil
\end{figure}

To answer these questions, various of the existing general-lossless compression algorithms  have been  directly applied on several compressed streams generated from our recent compression model \cite{cp:abuadbba2017gaussian}.  Fig \ref{fig:singlestreamCR}  shows the exact Compression Ratio (CR) of single  compressed streams after applying many of current lossless algorithms from both entropy and dictionary fields such as Huffman \cite{cp:less:Huff:zhang2009new},  Gaussian based Arithmetic coding \cite{cp:less:ac:bzip2:kraus2009loooseless} and Lempel-Ziv \cite{cp:less:lempelziv:gerek2008}. The CR varies between  $0.5 -$to$- 1.3$ with an average of  $0.9$. CR $\le 1$  means increasing the size rather than decreasing.  Therefore, the observation from the experiment (A) is that it is ineffective to re-compress an already compressed single stream alone using general compressors.

\begin{figure}[!h] %proposedarchiticture
	\centerline
	{\includegraphics[scale=0.40]{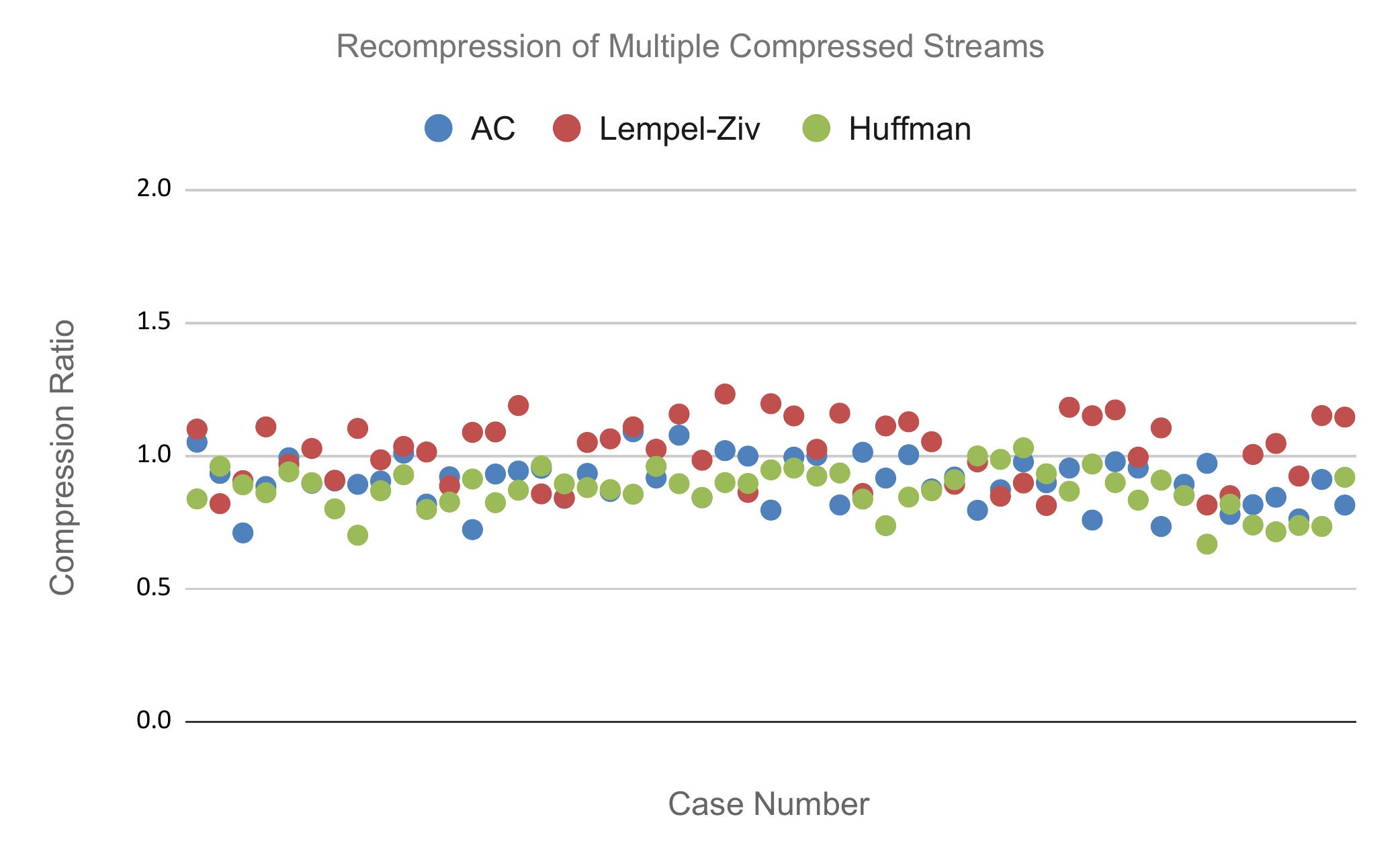}}
	\caption{Average compression ratio after re-compressing the multi-incoming compressed streams all together which was very poor (CR is $\le 1$).} 
	\label{fig:AVeragesinglestreamCR}
	\hfil
\end{figure}

Similarity, we repeated the same re-compression experiments but on collective (i.e. all together) multi-incoming compressed streams. We set  56 streams per experiment. Fig. \ref{fig:AVeragesinglestreamCR}) shows the CR of these experiments, which varies between $0.5 -$to$- 1.1$ with an average of $0.7$. This  is worse than the single re-compression. Hence, the observation from the experiment  (B) is that it would also be discouraged to re-compress a collective of compressed streams directly together. 

% \vspace{0.2cm}
% \noindent{\bf Summary:} It is evident  from the experiments that it is ineffective to re-compress the multi incoming compressed streams by directly  applying current lossless algorithms. This is because the redundancy in these streams is already exploited and re-applying the same algorithms will have little effect if not worse. Also, imposing all compressed streams together is not useful due to the vast dissimilarity in their features. 
% In other words, forcing all together will boost the noise and so decrease the chances of size reduction. 
\begin{figure*}[!t] %proposedarchiticture
	\centerline
	{\includegraphics[scale=0.30]{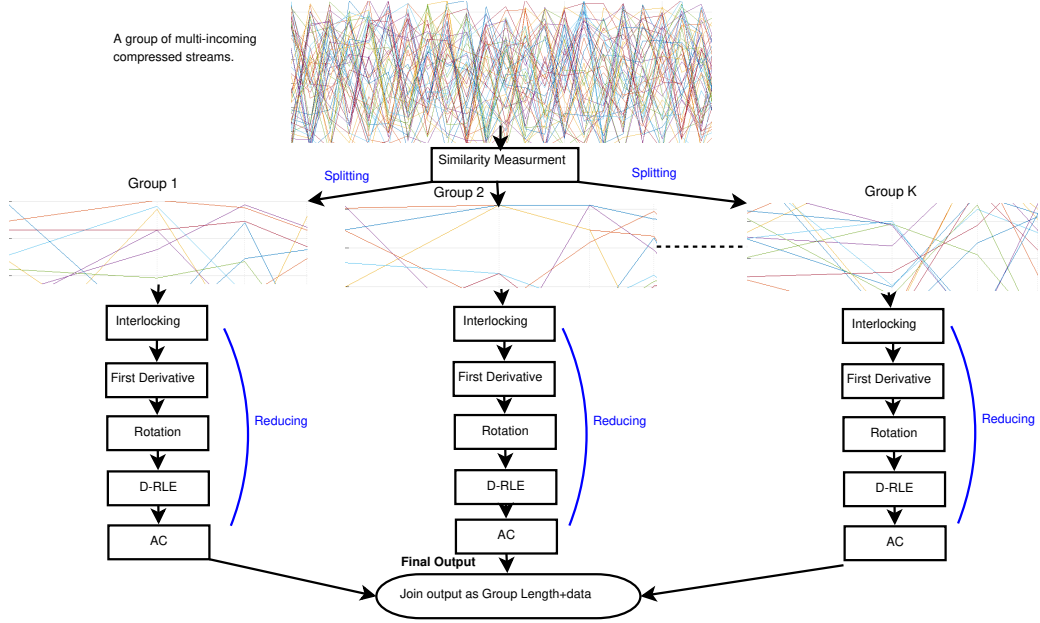}}
	\caption{An overview of the steps undertaking in our model where similarity measurement technique used to split the compressed streams. Then, the aggregation and size reduction steps are performed.} 
	\label{fig:modelflow}
	\hfil
\end{figure*}
\section{Our Re-compression Model}\label{sec:methodology}
A possible way to reduce the noise in the collective compressed streams is to explore the potential similarities among subgroups of these compressed streams and design an interlocking friendly compression algorithm. Therefore,  the unsupervised learning technique i.e., K-Means clustering has been used as a similarity measurement to classify the compressed streams into subsets. The streams in every subset have been interlocked  followed by the first derivative to reduce the values' space and increase the redundancy. After that, both BWT and MTF have been applied to rotate and rearrange the readings in a more consecutive format before employing the developed dynamic RLE. Finally, entropy coding is performed. Fig. \ref{fig:modelflow} demonstrates an overview of our proposed model.

\subsection{Similarity Measurement - K-Means}
% Based on our preliminary results, mixing all compressed streams will result in a poor re-compression ratio due to the dissimilarity in streams characteristics which increase the noise. Therefore, 

K-means \cite{kmeans:lloyd1982least} is used as a similarity measurement to cluster $n$ observations into $K$ groups and avoid the noise of mixing all together.  The main idea is that, let us assume $(x_1,x_2,...,x_n)$ are the $n$ incoming compressed streams where each stream is a d-dimensional vector. K-means will partitions the $n$ streams into $K(\leq n)$ groups $(G_1,G_2,..,G_k)$ by summation of distance functions of each point in the group to $K$ center. The objective is depicted in Eq \ref{eq:kmeans}

\begin{equation}\label{eq:kmeans}
\sum_{i=1}^{k}\sum_{j=1}^{\sigma_i}(\left \| x_i - \mu _j \right \|)^2
\end{equation}

where $\sigma_i$ is the number of data points in the $i^{th}$ group, $\mu _j$ is the center of the $i^{th}$ group and $\left \| x_i - \mu _j \right \|$ is the Euclidean distance between $x_i $ and $ \mu _j$.

The algorithm started by selecting cluster centers $\mu_j$. The distance between each reading point $x_i$ and $\mu_j$ is then calculated. Next, the reading point $x_i$ is assigned to $\mu_j$ based on the best minimum distance. After that, a new cluster center $\mu_j$ is recalculated as shown in Eq. \ref{eq:kmeansCen}

\begin{equation}\label{eq:kmeansCen}
\mu_i=\left ( \frac{1}{\sigma_i}\right ) \sum_{j=1}^{\sigma_i} x_i
\end{equation}
where $\sigma_i$ represents the number of reading points in $i^{th}$ cluster. The distance between $x_i$ and $\mu$ is then recalculated.  The assignment process will be repeated (See Eq \ref{eq:kmeansassign}), until no further data points need to be reassigned.  

\begin{equation}\label{eq:kmeansassign}
G_i=\left \{  x_p: \left \| x_p - m_i \right \|^2 \leq  \left \| x_p - m_j \right \|^2 \forall _j,1\leq j\leq k \right \}
\end{equation} 
 
The most crucial part is how centroid points are chosen. Therefore, to avoid the exponential time complexity of the standard algorithm, the idea proposed by Arthur and Vassilvitskii \cite{kmeans:arthur2007k}  has been used by utilizing a heuristic to find the centroid seeds for the algorithm as follows. Only one random center $\mu$ is uniformly chosen from among the readings. Then, the distance between $x_i$ and the closest center (i.e. chosen one) is computed. Next, one of the readings is chosen to be the new center $\mu$ using a weighted distribution probability (See Eq \ref{eq:kmeans++}). These steps are repeated until k centers are chosen.
\begin{equation}\label{eq:kmeans++}
\frac{d^2 \left ( x_i,\mu_p \right )}{\sum_{\left \{ j,x_j\in \Re_p \right \}} \left ( x_j,\mu_p \right )}
\end{equation}

where $\Re_p$ is the group of all observations nearest to centroid. $\mu_p$ and $x_i$  belong to $\Re_p$.

\subsection{Size Reduction}
After completing the similarity measurement process, each resultant group is combined and its size will be reduced using the following steps.  
\subsubsection{Readings Interlocking}
Let us assume $G_i$ is one of the resultant groups. It has $1,2,...,n$ vectors that represent multiple  compressed streams as shown in Eq \ref{eq:interlock}. 
\begin{equation}\label{eq:interlock}
G_{m,n} = 
\begin{pmatrix}
a_{1,1} & a_{1,2} & \cdots & a_{1,n} \\
b_{2,1} & b_{2,2} & \cdots & b_{2,n} \\
\vdots  & \vdots  & \ddots & \vdots  \\
c_{m,1} & c_{m,2} & \cdots & c_{m,n} 
\end{pmatrix}
\end{equation}
Therefore, these streams will be overlapped to exploit  similar features (See Eq \ref{eq:overlap}).
\begin{equation}\label{eq:overlap}
a(1,1),b(2,1),...,c(m,1)
\end{equation} 

To avoid any sharp exponential deviations in the overlapped readings and  increase the redundancy, the first derivative is applied as shown Eq \ref{eq:derivative}.
\begin{equation}\label{eq:derivative}
f_d = [\Upsilon (2)-\Upsilon(1) \Upsilon(3)- \Upsilon()2)...\Upsilon(\Lambda)-\Upsilon(\Lambda-1)]  
\end{equation} 

where $\Upsilon$ represents data points in the combined stream and $\Lambda$ is the latest value.

\subsubsection{Burrow Wheeler Transform (BWT)}
Based on experimental observations, the resultant values after applying the first derivative reflects that there are high redundancies but in a very scattered format which impedes  any size-reduction attempts. Consequently, a well-known transformation technique called BWT is employed to reshuffle the samples which result in a long consecutive and identical sequence. Originally, BWT has been proposed by Michael Burrows and David Wheeler \cite{bwt:bw1994} to rearrange text streams into a format that boosts its  compressibility by utilizing mechanisms such as MTF and RLE. The advantage of this algorithm is that zero additional overhead needed to reverse it. Basically, the data (i.e. 1 to $n$) is rotated lexicographically. Let us assume $\Omega$ be   textual or numerical of symbols group form (i.e.  numerical in our algorithm)  to be compressed.

\begin{equation}\label{eq:bwtmain}
\Omega =\Omega_1,\Omega_2,...,\Omega_n
\end{equation}

Iteratively, the vector $\Omega$ is rotated to the left which results in a new 2D matrix called $\beta $, as shown in Eq. \ref{eq:bwtbeta}

\begin{equation}\label{eq:bwtbeta}
\beta  = 
\begin{pmatrix}
\Omega_1  & \Omega_2 & \Omega_3&\cdots & \Omega_n \\
\Omega_2  & \Omega_3 &\cdots & \Omega_n& \Omega_1 \\
\Omega_3   &\cdots & \Omega_n& \Omega_1 & \Omega_2\\
\vdots  & \vdots  & &\ddots & \vdots  \\
\Omega_n  & \Omega_1 & \Omega_2&\cdots & \Omega_n-1 
\end{pmatrix}
\end{equation}

From Eq. \ref{eq:bwtbeta}, it is obvious that each rotation of $\Omega$ is represented as a row in $\beta$. These rows are then sorted in ascending order which will generate a new version of the matrix called $\widetilde{\beta}$. Only the last column $C$ of $\beta$ and the original block index $I$ are kept to be used for retrieving the original order. For clear understanding, a portion of the resultant first derivative values are presented in Table \ref{tb:conversion}. These samples have been replaced by characters (i.e. represent $\Omega$) for the sake of simplicity. $\beta$ is generated by rotating $\Omega$ for $n$ (i.e. the number of elements) times. 
%as shown in Eq. \ref{eq:bwtbetaeg}
% \begin{equation}\label{eq:bwtbetaeg}
% \beta =
% \begin{Bmatrix}
% \color{blue}\$ & a & b & a & a &b & a  \\ 
% a  & \color{blue}\$  & a & b & a & a &b \\ 
% b & a & \color{blue}\$  & a & b & a & a \\ 
% a & b & a  & \color{blue}\$  & a & b & a \\ 
% a & a & b & a  & \color{blue}\$  & a & b \\ 
% b &  a & a & b & a  & \color{blue}\$  & a  \\ 
% a &b &  a & a & b & a  & \color{blue}\$ 
% \end{Bmatrix}
% \end{equation}
%where $\$$ represents the start of data. 
The rows of $\beta$  will then be sorted which results in a new form $\widetilde{\beta}$ as depicted in Eq. \ref{eq:bwtwildbetae.g}

\begin{equation}\label{eq:bwtwildbetae.g}
\widetilde{\beta} =
\begin{Bmatrix}
a  & \color{blue}\$  & a & b & a & a &\color{red}\boldsymbol{b} \\ 
a & a & b & a  & \color{blue}\$  & a & \color{red}\boldsymbol{b} \\ 
\color{red}\boldsymbol{a} &\color{red}\boldsymbol{b} &  \color{red}\boldsymbol{a} & \color{red}\boldsymbol{a} & \color{red}\boldsymbol{b} & \color{red}\boldsymbol{a}  & \color{red}\boldsymbol{\$} \\
a & b & a  & \color{blue}\$  & a & b & \color{red}\boldsymbol{a} \\ 
b &  a & a & b & a  & \color{blue}\$  & \color{red}\boldsymbol{a} \\ 
b & a & \color{blue}\$  & a & b & a & \color{red}\boldsymbol{a}\\ 
\color{blue}\$ & a & b & a & a &b & \color{red}\boldsymbol{a}   
\end{Bmatrix}
\end{equation}

The last column $C$ and the index I (e.g. 3 in this example) represent the output and used by the decoder to recover the original form by inversing the above steps.

\begin{table}[]
	\centering
	\caption{Conversion from Numeric to characters}
	\label{tb:conversion}
	\begin{tabular}{l|l|l|l|l|l|l}
		\hline \hline
		$\omega$&	309& 501& 309&  309& 501& 309  \\\hline
		$Ch$&	a& b& a&  a& b& a \\\hline\hline
	\end{tabular}
\end{table}

%This is achieved by building a temporary $n\times n$ matrix  (i.e. $n$ is the elements number) and $C$ represents its last column. By sorting this column, the first column is retrieved. Next, all successive pairs are recovered using those two columns and the resultant matrix is similar to $\widetilde{\beta}$. Subsequently, by using the index $I$ retrieving the original form will be easy.  

\subsubsection{Move-To-Front (MTF)}
Despite  the resultant BWT values precisely gathers identical symbols in long runs, these values still  sharply vary from very low (e.g. 20 and 21) to much higher numbers (e.g. 4000 and 6000). Consequently, MTF transform is employed to boost the influence of any entropy based encoder (e.g. Arithmetic Coding) to achieve the highest compression rate.  MTF is a lightweight mechanism introduced by  Ryabko \cite{mtv:mtf1980} to enhance the low values (e.g. close to zero) probability while minimizing the high values in a given list of data. The basic idea is that the data list symbols are substituted by their positions in a unique list. With this, the long sequential identical symbols will be substituted by as many zeros, whereas a posterior (i.e. not regularly used) symbols will be exchanged by larger values.     
\begin{table}[!b]
	\centering
	\caption{MTF of $L = \left [ b,b,a,a,a,a,a,a \right ]$ and $u = \left [ a,b \right ]$ }
	\label{tb:mtf}
	\begin{tabular}{l|c|c|c|c|c|c|c|c}
		\hline\hline
		$L_i$& b &b &a &a &a &a &a &a \\\hline
		  $u$&a, b  &b, a&b, a&a, b&a, b&a, b&a, b& a, b \\\hline
	$\vartheta$&1   &0  &1  &0  &0  &0  &0  &0\\\hline
	
	\end{tabular}
\end{table}
% Let us assume $L$ as an obtained list from BWT process and $u$ is the unique symbols of $L$ (i.e. $u\in L$). Then, the summary of MTF process can be depicted in the following steps. (a) $L$ is used to populate $u$. (b) Every item $L_i$ of vector $L$ is substituted as its the symbol numbers preceding it in $u$. (c) Latter output is built as a list $\vartheta$ by collecting the resultant codes of step 2. The recovery process is the opposite of  these steps.  
Let us assume the BWT resultant list is $L = \left [ b,b,a,a,a,a,a,a \right ]$ and so  its unique list is $u = \left [ a,b \right ]$ (See Table \ref{tb:mtf}). The initial token $L_0$ is $b$, and its preceded by one symbol in $u$. Therefore, the digit one is produced in $\vartheta$ and the symbol $b$ is moved to the front of $u = \left [ b,a \right ]$. The next token $L_1$ is $b$ which is the first in $u$, and so the produced value is $0$ and no need to update  $u$. These steps are continued until the last token is reached the resultant output will look like $\vartheta = \left [ 1,0,1,0,0,0,0,0 \right ]$. 

\subsubsection{Run Length} \label{sec:RLE}
The MTF output includes  series of  identical sequential tokens. Consequently, to exploit this fact, a simple mechanism called run length (RLE) \cite{comp:Salmon2004} is employed before the entropy encoding. RLE focuses on substituting the similar consecutive symbols by their count. Let  $s$ represents a symbol that appears $n$ sequential times in a vector $V$.  The $n$ cases are then substituted by $ns$. The $n$  consecutive appearances of the symbol are called run length. For instance, the sequential zeros in $\vartheta = \left [ 1,0,1,0,0,0,0,0 \right ]$ will be $\vartheta = \left [ 1,0,1,0\#5 \right ]$. However, the observation was that a naive implanting of RLE is not always useful and may increase rather than decrease. This is due to its static nature where each consecutive symbols are replaced even in the case of no repeated tokens occurs. RLE is improved to be a dynamic (by adding 1 bit) based on thresholds  $t=t1,t2,..,t_n$ that monitor the consecutive occurrences of the symbols. %Only one bit is added at the beginning of the resultant encoded stream to indicate whether RLE encoding process was employed or not.   
\begin{table}[!b]
	\centering
	\caption{The message $m$ probability Distribution }
	\label{tb:frefprob}
	\begin{tabular}{c|c|c}
		\hline \hline
		Symbol&	Probability&Accumulative range \\\hline
		a&	0.2& (0.0,0.2) \\\hline
		b&	0.3& (0.2,0.5)\\\hline
		c&	0.1& (0.5,0.6) \\\hline
		d&	0.2& (0.6,0.8) \\\hline
		e&	0.1& (0.8,0.9) \\\hline
		f&	0.1& (0.9,1.0) \\\hline \hline
	\end{tabular}
\end{table}
\begin{figure}[!h] %proposedarchiticture
	\centerline
	{\includegraphics[scale=0.25]{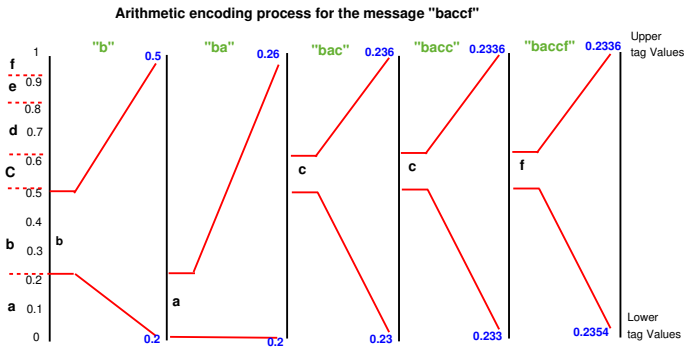}}
	\caption{Graphical representation for Arithmetic encoding steps for a message $b,a,c,c,f$} 
	\label{fig:arethmaticencode}
	\hfil
\end{figure}
\subsubsection{Arithmetic Coding}
An entropy coding mechanism called Arithmetic Coding (AC) is ultimately employed in our model to achieve the highest possible compression ratio. AC is a widely-known variable length statistical coding by which repeatedly occurred values are represented with fewer bits whereas less frequently appearing tokens are symbolized with higher bits number \cite{comp:Salmon2004}. AC proved its superiority in most respects to other well-known entropy algorithms such as Huffman coding. This is due to its succinct representation of the entire message in a single value as a fraction $n$ where  $\left (0.0 \leq  n < 1.0  \right )$, whereas other algorithms working on separating the input into isolated component tokens and substituting each with a unique code.

The general idea is that after choosing a specific interval, the symbols list will be scanned and based on its tokens probabilities, the ultimate interval will be narrowed.

% AC core steps are summarized as follows.  
% \begin{enumerate}
% 	\item   The current interval is specified as (0,1).
% 	\item The following two steps are repeated for each token $S_i$ in the data vector.
% \begin{enumerate}
% 	\item The current  interval  is divided into subintervals under the condition that their sizes are proportional to the tokens  probabilities.  
% 	\item A subinterval for $S_i$ is chosen which will represent the new current interval.
% \end{enumerate}
% 	\item After processing the entire data vector, the result should be any value that distinctly identifies the present interval (i.e. any value in the current interval). 
% \end{enumerate}

% Interestingly, the deeper scanned values in the data vectors are, the smaller current interval is obtained. The resultant output is a single measure called tag value that does not include the individual tokens codes.

To demonstrate AC mechanism, let us assume that an entire  message $M$ that has a probability distribution as given in Table \ref{tb:frefprob}. For brevity, a fraction of that message $\tilde{m}=(b,a,c,c,f)$ is encoded (see Fig. \ref{fig:arethmaticencode}). The probability boundary is between $(0,1)$. To begin with, due to the occurrence of symbol $b'$, the tag value should be in the range $(0.2,0.5)$. After that, the token $'a'$  is appeared, so the present interval between $(0,0.2)$ which will be used to calculate the lower boundary: $w_n=l_{n-1}+(w_{n-1}-\rho_{n-1})\times F_x(x_{n-1})$ and upper boundary: $\rho_n=w_{n-1}+(\rho_{n-1}-\rho_{n-1})\times F_x(x)$. $F_x$ is the frequency accumulation.

% appeared in the equations \ref{eq:lowlimit} and \ref{eq:uplimit}. 

% \begin{equation}

% \label{eq:lowlimit}
% \end{equation}
% \begin{equation}

% \label{eq:uplimit}
% \end{equation}
The resultant tag values of symbols  sequence 'ba' are  $(0.2,0.26)$. This will  continue for the full message in an accumulative manner. The ultimate tag values output has been summed up in Fig. \ref{fig:arethmaticencode}. The average of both the final upper and lower tags $\frac{w_n+\rho_n}{2} =\frac{0.23354+0.2336}{2}=0.23357$ represents the compressed value and will be transformed into binary.

The decoder side requires both the average value and the message probabilities. Subsequently, it proceeds through similar steps but in an inverse manner where the probability accumulation is used to find the symbols.

\section{Decompression and Recovery}

The recovery process is almost similar to the steps stated above but in the opposite manner. It begins by Arithmetic decoding followed by RLE if needed (i.e. based on the conditions mentioned in Section \ref{sec:RLE}). Then, MTV and BWT are applied respectively. The output represents the derivative values and so their inverse process is employed to reconstruct the actual symbols. These symbols are the compressed streams in an interlocking way. From that, they are disunited to  their single compressed streams and so their original format is recovered in a lossless format. 
\section{Evaluation}\label{sec:evaluation}
Various matrices are used to examine the  effectiveness of our lossless size reduction model. 
\begin{figure}[!h]
	\centerline
	{\includegraphics[scale=0.22]{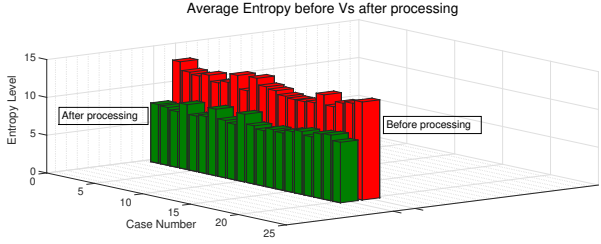}}
	\caption{Comparison between the average entropy calculated from  the compressed streams before and after aggregation and processing.} 
	\label{fig:entropyfig}
\end{figure}
\subsection{Silhouette Measurement}
To validate the coherence  of the used similarity measurement clustering techniques, a mathematical model called Silhouette is used. It is a graphical representation technique  proposed by Peter J. Rousseeuw \cite{silh:rousseeuw1987silhouettes} that proves the consistency within the data cluster and clearly reflects the correlation of the objects within that group. The Silhouette model produces a value in the range of $-1$ to $1$, in which the higher the value, the more the object is well matched to that group and vice versa. Let us assume $n$ vectors are clustered into $K$ groups using a similarity measurement model (e.g. $K$-means). For each group $G$, $x(i)$ represents the average dissimilarity (i.e. distance ) of $i$ within the group (i.e. the least the value, the better the matching). Also, let $y(i)$ be the lowest dissimilarity of $i$ within the group. With this, silhouette $s$ can be defined as follows.

\begin{equation}\label{eq:silhoute}
s(i)=\frac{y(i)-x(i)}{\max\{x(i),y(i)\}}
\end{equation}

\begin{equation}\label{eq:silhoute2}
s(i)=\begin{cases}
1-\frac{x(i)}{y(i)}, & \text{ if  } x(i)<y(i) \\ 
0, & \text{ if } x(i)=y(i) \\ 
\frac{x(i)}{y(i)}-1 & \text{ if } x(i)>y(i)
\end{cases}
\end{equation}
From \ref{eq:silhoute} and \ref{eq:silhoute2}, it can be derived that $-1\leq s(i)\leq 1$. 

% \begin{figure*}[!t] %proposedarchiticture
% 	\centerline
% 	{\includegraphics[scale=0.4]{silhouetteme_kmeans.eps}}
% 	\caption{Graphical representation of the consistency among various compressed streams combined (i.e. K-means) clusters. Obviously, six and eight clusters are the best in terms of the cohesion.   } 
% 	\label{fig:silhouete_kmeans}
% 	\hfil
% \end{figure*} 
\begin{figure*}[!h] %proposedarchiticture
	\centerline
	{\includegraphics[scale=0.35]{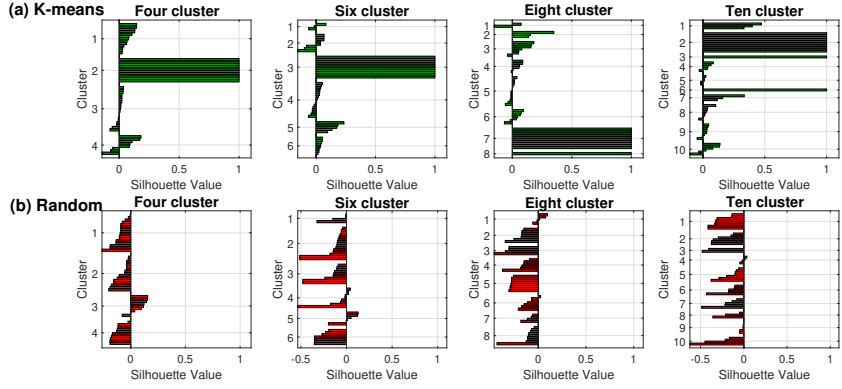}}
	\caption{Comparison between the consistency among the clusters by changing two parameters. These are the number of clusters and the similarity measurement technique (i.e. K-means vs Rand). } 
	\label{fig:silhoueteKmeansVSrand}
	\hfil
\end{figure*}

\subsection{Theoretical Entropy}
The entropy of a signal in the information theory field represents the lowest bit-rate (i.e. the optimum compression assumed) needed for transmitting this signal \cite{comp:shannon1948}. Consequently, to monitor the influence of  preprocessing  the compressed streams in our model, the theoretical entropy is measured for every smart meter compressed stream before and after employing our model. After that, the quantitative calculation comparison is performed between the theoretical entropy and the accomplished size reduction ratio.
Let us assume a compressed incoming stream  consisting of the symbol points  $d[1], d[2], ..., d[N]$. The optimum likelihood entropy (i.e. in bits) is calculated as 

\begin{equation}
H(d) \triangleq -\sum_{i\in \mathbb{R}(d)} \widehat{p}(i) \log_2 (\widehat{p}(i))
\label{eq:entropy}
\end{equation}
\begin{equation}
\widehat{p}(i) \triangleq \frac{1}{N} \sum_{n=1}^{N} \delta_i(d[n])
\label{eq:probentropy}
\end{equation}
\begin{equation}
\delta_i(d[n]) \triangleq \begin{cases}
1, & \text{ if } d[n]=i \\ 
0, & else
\end{cases}
\label{eq:deltaentropy}
\end{equation}

where $\widehat{p}(i)$ is the experimental probability of $i\in \mathbb{R}$ and $\mathbb{R}(d)$ represents the range of $d$. The smallest entropy (i.e. optimum case) happens when all $d$ symbols are equal, which results in $H_{min}=-1 \log_2 (1)=0$.

On the other hand, the worst-case (i.e. maximum entropy) happens when each
symbol in $\mathbb{R}$ occurs at the somehow similar frequency $1/\mathbb{R}$, in which
$|\mathbb{R}|$ reflects the original elements in  $\mathbb{R}$  (See Eq. \ref{eq:highestH}).

\begin{equation}
H_{\max}=-\sum_{i\in \mathbb{R}} \frac{1}{|\mathbb{R}|} \log_2 (\frac{1}{|\mathbb{R}|})=\log_2 |\mathbb{R}|
\label{eq:highestH}.
\end{equation}

\section{Experimental Ratio}

The compression ratio (CR) is the essential benchmark to  measure any proposed compression model empirically. Let us symbolize the original compressed streams block $O$ (i.e. it's unit in bit or byte) and the resultant re-compressed symbols $R$. Consequently, the experimental CR in the results section is measured as $CR=\frac{O}{R}$. The widely-known leading power quality storage standard for electric waveforms power system  utilized in most of smart meters called PQDIF (Power Quality Data Interchange Format) has been employed in producing the multi compressed streams dataset.  Every Reading represented as 16 bit and the typical suggested block size is  used which is about 1500 readings. The entropy-based compression model published in \cite{cp:abuadbba2017gaussian} is employed which proven to give the optimum lossless compression ratio $3.8:1$. 
\begin{table*}	
	\caption{Compression ratio}	
	\centering
	\begin{center}
		\begin{tabular}{c|c|c|c|c|>{\columncolor[gray]{0.9}}c}
			\hline \hline
		Record&	Cluster No &Static RLE (Rand) &Static RLE(Kmeans) &	Dynamic RLE (Rand) &Dynamic RLE (Kmeans)  \\\hline
		1& 2	&1.03	&0.40	&1.30	&1.75\\
		2& 2	&1.08	&0.31	&1.36	&1.73\\
		3& 2	&1.12	&1.46	&1.31	&1.77\\
		4& 2	&1.01	&1.52	&1.30	&1.73\\\hline
		5& 4	&1.07	&1.53	&.35	&1.81\\
		6& 4	&1.10	&1.38	&1.30	&1.76\\
		7& 4	&1.18	&1.53	&1.33	&1.81\\
		8& 4	&1.12	&1.63	&1.31	&1.83\\\hline
		9& 6	&1.08	&1.59	&1.39	&1.90\\
		10& 6	&1.12	&1.43	&1.33	&1.90\\
		11& 6	&1.20	&1.55	&1.35	&1.98\\
		12& 6	&1.17	&1.67	&1.33	&1.98\\\hline
		13& 8	&1.05	&1.69	&1.40	&2.11\\
		14& 8	&1.16	&1.55	&1.44	&2.13\\
		15& 8	&1.23	&1.67	&1.35	&2.10\\
		16& 8	&1.29	&1.77	&1.42	&2.19\\\hline
		17& 10	&1.10	&1.63	&1.36	&1.94\\
		18& 10	&1.15	&1.43	&1.40	&1.95\\
		19& 10	&1.21	&1.60	&1.38	&2.04\\
		20& 10	&1.19	&1.70	&1.40	&2.03\\\hline
		21& 12	&1.04	&1.51	&1.31	&1.85\\
		22& 12	&1.11	&1.41	&1.33	&1.89\\
		23& 12	&1.20	&1.58	&1.36	&1.97\\
		24& 12	&1.10	&1.66	&1.42	&2.04\\\hline
		\rowcolor[gray]{0.9}
		Average	& &1.13	&1.55	&1.36	&1.92 \\\hline \hline
			
		\end{tabular}
	\end{center}
	\label{tb:results}
\end{table*}
\section{Implementations} \label{sec:implementation}
\subsection{Datasets}\label{subsec:dadaset}
\begin{figure}[!b] %proposedarchiticture
	\centerline
	{\includegraphics[scale=0.25]{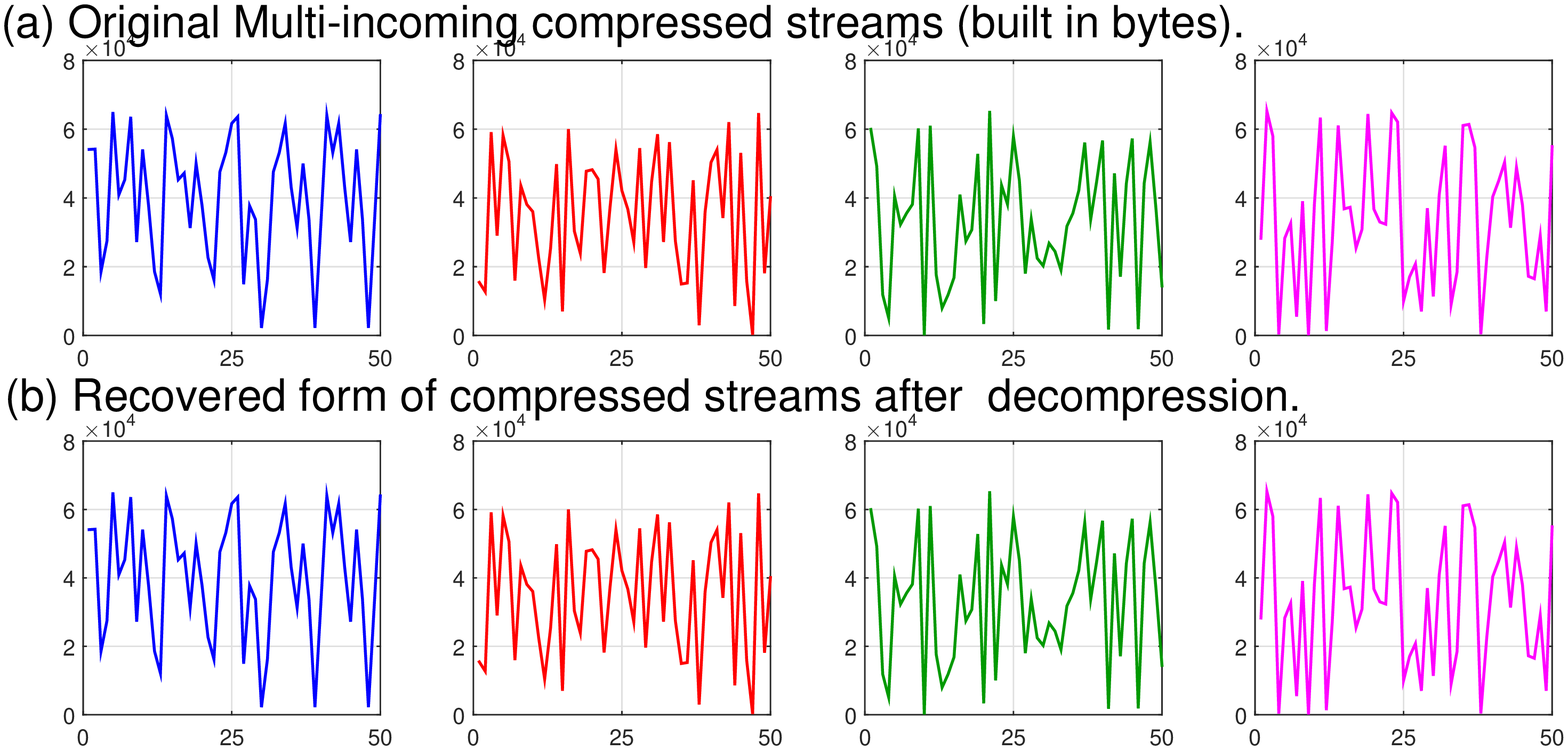}}
	\caption{4 examples of compressed watts consumptions’ readings collected from different homes: (a) Direct plot of single compressed streams form, and (b) Plot of  these streams after dis-aggregation and recovery.} 
	\label{fig:streamssample}
	\hfil
\end{figure} 
% \begin{figure*}[!t] %proposedarchiticture
% 	\centerline
% 	{\includegraphics[scale=0.5]{3Dbar_groups.eps}}
% 	\caption{4 groups of achieved re-compression ratio of multi-incoming compressed streams by changing the similarity measurement technique, number of clusters and RLE. Every group contains 56 compressed streams. } 
% 	\label{fig:3Dbar_groups}
% 	\hfil
% \end{figure*} 
The Laboratory for Advanced System Software collected and published a detailed smart meters datasets as a part of a project called "Smart" \cite{dataset:smartproject2012,dataset:barker2012smart}. These datasets have been thoroughly used in our experiments. The datasets entries represent a periodical (i.e. per minutes) readings from three houses for more than three months. 
% The entries can be classified into (i) power consumption as watts and heat-index, and (ii) environmental features as in/out temperature, in/out humidity and wind-chill. 
Also, a detailed electricity consumption (i.e. per minutes) from about 400 anonymous premises for $(24\times 30\times 3) hrs$ is provided. 
% According to the spatial and temporal aggregations definition,  these symbols are temporal due to their separate periodical collection from every individual premise by equipping it with a smart meter.
Our recent Gaussian based compression \cite{cp:abuadbba2017gaussian} is applied on every single stream, to generate the multi-incoming compressed streams as explained in the evaluation Section  \ref{sec:evaluation}. The compressed streams are laid out as the bed-test  in  all  of our experiments.

\subsection{Experiments and Results}
Our prime executed experiments will be done at the operation centers or cloud level after receiving an overwhelming amount of compressed streams from a huge number of premises. The experiments can be categorized into  (a) similarity measurements, interleaving and size reduction processes, and (b) an original format recovery. Both categories are designed in such a way that can be done in parallel to take advantage of cloud power. To obtain unbiased outcomes, all compressed streams  mentioned above have been employed in our model. 
%Identically, many well-known  lossless compression algorithms have been applied directly on the multi-compressed streams as shown in Fig \ref{fig:singlestreamCR} and \ref{fig:AVeragesinglestreamCR} to accurately provide a clear comparison and prove that these streams can not be re-compressed using existing models.

\begin{figure}[!t] %proposedarchiticture
	\centerline
	{\includegraphics[scale=0.25]{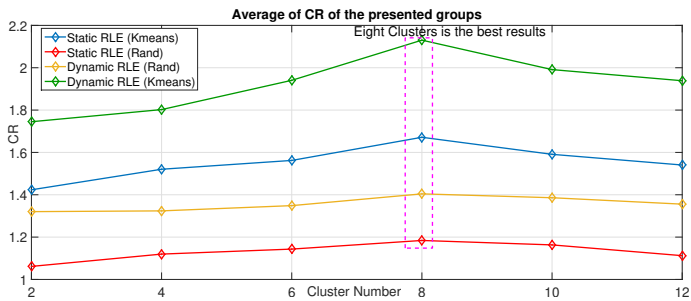}}
	\caption{The average of re-compression ratio of the four groups examined in Table \ref{tb:results} which shows the best combination of our technique parameters.} 
	\label{fig:AVeragegroupstreamCR}
	\hfil
\end{figure}

For brevity in this paper, the results have been summarized as follows.  
% Fig \ref{fig:silhouete_kmeans} highlights the examination process of our similarity measurement  technique (i.e. K-means) to select the best $K$ number (i.e. Without loss of generality, 6 and 8 are the best in the case of 56 streams) using Silhouette benchmark. 
Fig. \ref{fig:silhoueteKmeansVSrand} reflects the superiority (more group cohesiveness $>$0) as a graphical comparison (i.e. using Silhouette benchmark) between our similarity measurement using $K-means$ against a random agnostic  grouping. 
Secondly, Fig. \ref{fig:entropyfig} shows a comparison   between  the   average   entropy   calculated   from   the compressed streams of few random streams together vs a processed cluster after using our model. The entropy has been noticeably reduced from around 13 bits into 7 bits.  
% Fourthly, Fig \ref{fig:3Dbar_groups} emphasizes the possible enhancement  on CR level by changing various parameters in our algorithm such as static RLE, dynamic RLE, agnostic similarity measurement, K-means similarity measurement in relation to the number of clusters. Each of these groups contains the same number of compressed streams. 
Thirdly, Table \ref{tb:results} presents the exact CR achieved from the 4 configurations starting from Random Cluster and Static RLE to Kmeans cluster with Dynamic RLE. This proves the feasibility of improving CR using clustering and dynamic RLE. Fourthly, Fig \ref{fig:AVeragegroupstreamCR} shows the average of CR ratio obtained from the above four groups (i.e. 8 Kmeans clusters + dynamic RLE is the optimum). Finally, Fig \ref{fig:streamssample} shows an example of a plot of various original compressed streams before and after the aggregation and size reduction process.

\subsection{Discussion} 
%It is obvious from the early experiments that it is ineffective to re-compress the multi incoming compressed streams by simply applying current lossless algorithms. This is because the redundancy in these streams is already exploited and re-applying the same algorithms will have little effect if not worse (Revisit Fig \ref{fig:singlestreamCR} and \ref{fig:AVeragesinglestreamCR} ). On the other hand, imposing all compressed streams together is also not useful due to the huge dissimilarity in their features. In other words, forcing all together will boost the noise and so decrease the chances of size reduction. 
Recall, the main question drives this work: Can  the  multi-incoming  smart  meter  compressed streams be re-compressed? By exploring the potential similarities in various compressed streams using a similarity measurement techniques, our model demonstrated that it is possible to re-compress multi-incoming compressed streams.  We achieved up to 2:1 size reduction level in the optimum setting and 1.9:1 in average. This means every 1Gigabyte byte can be reduced to $	\sim$500 Megabyte. This has been emphasized theoretically by comparing the entropy before and after our technique (See Fig. \ref{fig:entropyfig}) and experimentally as presented in Table \ref{tb:results}. The scalability impact  in the number of compressed streams that goes into similarity measurement  along with the number of samples per stream before the initial compression  remain to be addressed in our future work. 

\section{Conclusion}\label{sec:conclusion}
In this paper, a novel lossless re-compression algorithm was introduced to prove the possibility of reducing the size of the already compressed waveform smart meter readings. The main target was preprocessing the data to enhance the entropy.
This is successfully achieved by employing K-Means clustering as similarity measurement to classify the compressed streams into subsets to reduce the effect of uncorrelated compressed streams. The tokens of every subset have been interlocked  followed by the first derivative to reduce the values' space and boost the redundancy. After that, two rotation steps have been applied to rearrange the symbols in a more consecutive format before employing dynamic RLE. Finally, entropy coding is performed. Both mathematical and empirical experiments proved the possibility of enhancing the entropy (i.e. almost reduced by half) and the resultant size reduction (i.e. up to 50\%).

\bibliographystyle{unsrt}
\bibliography{References}

\begin{thebibliography}{10}

\bibitem{smartgrid:gungor2010opportunities}
Vehbi~C Gungor, Bin Lu, and Gerhard~P Hancke.
\newblock Opportunities and challenges of wireless sensor networks in smart
  grid.
\newblock {\em Industrial Electronics, IEEE Transactions on},
  57(10):3557--3564, 2010.

\bibitem{CIGREwebsite}
CIGRE.
\newblock D2-21 cigre working group d2.21, wg. broadband plc applications,
  http://www.cigre.org/what-is-cigre, accessed: 15.01.2020.

\bibitem{smartgrid:gungor2011smart}
Vehbi~C G{\"u}ng{\"o}r, Dilan Sahin, Taskin Kocak, Salih Erg{\"u}t, Concettina
  Buccella, Carlo Cecati, and Gerhard~P Hancke.
\newblock Smart grid technologies: communication technologies and standards.
\newblock {\em Industrial informatics, IEEE transactions on}, 7(4):529--539,
  2011.

\bibitem{cp:stateofart:tcheou2014}
Michel~P Tcheou, Lisandro Lovisolo, Mois{\'e}s~Vidal Ribeiro, Eduardo~AB
  da~Silva, Marco~AM Rodrigues, Joao Marcos~Travassos Romano, and Paulo~SR
  Diniz.
\newblock The compression of electric signal waveforms for smart grids: state
  of the art and future trends.
\newblock {\em Smart Grid, IEEE Transactions on}, 5(1):291--302, 2014.

\bibitem{cp:dwtsg:ning2011wavelet}
Jiaxin Ning, Jianhui Wang, Wenzhong Gao, and Cong Liu.
\newblock A wavelet-based data compression technique for smart grid.
\newblock {\em Smart Grid, IEEE Transactions on}, 2(1):212--218, 2011.

\bibitem{cp:liftingwp:tse2012real}
Norman~CF Tse, JohnY~C Chan, Wing-Hong Lau, Jone~TY Poon, and LL~Lai.
\newblock Real-time power-quality monitoring with hybrid sinusoidal and lifting
  wavelet compression algorithm.
\newblock {\em Power Delivery, IEEE Transactions on}, 27(4):1718--1726, 2012.

\bibitem{cp:damsin:tcheou2007optimum}
Michel~P Tcheou, Lisandro Lovisolo, Eduardo~AB da~Silva, Marco~AM Rodrigues,
  and Paulo~SR Diniz.
\newblock Optimum rate-distortion dictionary selection for compression of
  atomic decompositions of electric disturbance signals.
\newblock {\em Signal Processing Letters, IEEE}, 14(2):81--84, 2007.

\bibitem{cp:harmonic:ribeiro2007novel}
Mois{\'e}s~V Ribeiro, Seop~Hyeong Park, Jo{\~a}o Marcos~T Romano, and Sanjit~K
  Mitra.
\newblock A novel mdl-based compression method for power quality applications.
\newblock {\em Power Delivery, IEEE Transactions on}, 22(1):27--36, 2007.

\bibitem{stego:abuadbba2014wavelet}
Alsharif Abuadbba and Ibrahim Khalil.
\newblock Wavelet based steganographic technique to protect household
  confidential information and seal the transmitted smart grid readings.
\newblock {\em Information Systems (2014).}, 2014.

\bibitem{cp:less:singularval:2015desouza}
J.~C.~S. de~Souza, T.~M.~L. Assis, and B.~C. Pal.
\newblock Data compression in smart distribution systems via singular value
  decomposition.
\newblock {\em IEEE Transactions on Smart Grid}, (99):1--1, 2015.

\bibitem{cp:less:transgolomb:2016tale}
J.~E. Tate.
\newblock Preprocessing and golomb -rice encoding for lossless compression of
  phasor angle data.
\newblock {\em IEEE Transactions on Smart Grid}, 7(2):718--729, March 2016.

\bibitem{tripathi2018efficient}
Sharda Tripathi and Swades De.
\newblock An efficient data characterization and reduction scheme for smart
  metering infrastructure.
\newblock {\em IEEE Transactions on Industrial Informatics}, 14(10):4300--4308,
  2018.

\bibitem{cp:abuadbba2017gaussian}
Alsharif Abuadbba, Ibrahim Khalil, and Xinghuo Yu.
\newblock Gaussian approximation based lossless compression of smart meter
  readings.
\newblock {\em IEEE Transactions on Smart Grid}, 2017.

\bibitem{cp:dwt:santoso1997power}
Surya Santoso, Edward~J Powers, and WM~Grady.
\newblock Power quality disturbance data compression using wavelet transform
  methods.
\newblock {\em Power Delivery, IEEE Transactions on}, 12(3):1250--1257, 1997.

\bibitem{cp:bspline:meher2004integrated}
Saroj~K Meher, AK~Pradhan, and G~Panda.
\newblock An integrated data compression scheme for power quality events using
  spline wavelet and neural network.
\newblock {\em Electric power systems research}, 69(2):213--220, 2004.

\bibitem{cp:swt:panda2002data}
Ganapati Panda, PK~Dash, Ashok~Kumar Pradhan, and Saroj~K Meher.
\newblock Data compression of power quality events using the slantlet
  transform.
\newblock {\em Power Delivery, IEEE Transactions on}, 17(2):662--667, 2002.

\bibitem{cp:less:lempelziv:gerek2008}
{\"O}mer~Nezih Gerek and Dogan~G{\"o}khan Ece.
\newblock Compression of power quality event data using 2d representation.
\newblock {\em Electric Power Systems Research}, 78(6):1047--1052, 2008.

\bibitem{cp:less:ac:bzip2:kraus2009loooseless}
Jan Kraus, Tomas Tobiska, and Viktor Bubla.
\newblock Loooseless encodings and compression algorithms applied on power
  quality datasets.
\newblock In {\em Electricity Distribution-Part 1, 2009. CIRED 2009. 20th
  International Conference and Exhibition on}, pages 1--4. IET, 2009.

\bibitem{cp:less:Huff:zhang2009new}
Dahai Zhang, Yanqiu Bi, and Jianguo Zhao.
\newblock A new data compression algorithm for power quality online monitoring.
\newblock In {\em Sustainable Power Generation and Supply, 2009. SUPERGEN'09.
  International Conference on}, pages 1--4. IEEE, 2009.

\bibitem{cp:less:predictionlzma:kraus2012}
Jan Kraus, Pavel {\v{S}}t{\'e}p{\'a}n, and Leo{\v{s}} Kuka{\v{c}}ka.
\newblock Optimal data compression techniques for smart grid and power quality
  trend data.
\newblock In {\em Harmonics and Quality of Power (ICHQP), 2012 IEEE 15th
  International Conference on}, pages 707--712. IEEE, 2012.

\bibitem{kmeans:lloyd1982least}
Stuart Lloyd.
\newblock Least squares quantization in pcm.
\newblock {\em IEEE transactions on information theory}, 28(2):129--137, 1982.

\bibitem{kmeans:arthur2007k}
David Arthur and Sergei Vassilvitskii.
\newblock k-means++: The advantages of careful seeding.
\newblock In {\em Proceedings of the eighteenth annual ACM-SIAM symposium on
  Discrete algorithms}, pages 1027--1035. Society for Industrial and Applied
  Mathematics, 2007.

\bibitem{bwt:bw1994}
M.~Burrows and D.~J. Wheeler.
\newblock A block-sorting lossless data compression algorithm.
\newblock {\em Digital SRC Research Report}, (124), 1994.

\bibitem{mtv:mtf1980}
B~Ryabko.
\newblock Data compression by means of a "book stack”.
\newblock {\em Problems of Information Transmission}, 16(4):265--269, 1980.

\bibitem{comp:Salmon2004}
David Salomon.
\newblock {\em Data Compression: The Complete Reference}.
\newblock Springer-Verlag New York, 2004.

\bibitem{silh:rousseeuw1987silhouettes}
Peter~J Rousseeuw.
\newblock Silhouettes: a graphical aid to the interpretation and validation of
  cluster analysis.
\newblock {\em Journal of computational and applied mathematics}, 20:53--65,
  1987.

\bibitem{comp:shannon1948}
C.~E. Shannon.
\newblock A mathematical theory of communication.
\newblock {\em The Bell System Technical Journal}, 27(4):623--656, Oct 1948.

\bibitem{dataset:smartproject2012}
Sean Barker, Aditya Mishra, David Irwin, Emmanuel Cecchet, Prashant Shenoy, and
  Jeannie Albrecht.
\newblock Smart project.
\newblock \url{http://traces.cs.umass.edu/index.php/Smart/Smart}, 2012.

\bibitem{dataset:barker2012smart}
Sean Barker, Aditya Mishra, David Irwin, Emmanuel Cecchet, Prashant Shenoy, and
  Jeannie Albrecht.
\newblock Smart*: An open data set and tools for enabling research in
  sustainable homes.
\newblock {\em SustKDD, August}, 2012.

\end{thebibliography}

\end{document}